\documentclass[11pt]{article}          

\usepackage{graphicx}
\usepackage[dvips]{color}

\newcommand\Softitle[1]{\Large \bf \noindent \begin{center} #1
\end{center}\rm \normalsize \vskip.125in }%

\newcommand\Sofauthor[1]{\vskip.1in\noindent%
\large  \begin{center} \textsf{#1} \end{center}\rm \vskip-.2in}

\long\def\symbolfootnote[#1]#2{\begingroup%
\def\thefootnote{\fnsymbol{footnote}}\footnote[#1]{#2}\endgroup}

\let\title\Softitle
\let\author\Sofauthor

\let\address\Sofaddress

\begin{document}

\title{Does the speed of light depend upon the vacuum ?}

\author{Marcel Urban, Fran\c cois Couchot, Xavier Sarazin}
\address{LAL, Univ Paris-Sud, IN2P3/CNRS, Orsay, France}

\begin{abstract}
We propose a quantum model for the vacuum filled of virtual particle pairs. The main originality of this model is to define a density and a life-time of the virtual particles. Compared to the usual QED $(p,E)$ framework, we add here the $(x,t)$ space time parameters. 
We show how $\epsilon_0$ and $\mu_0$ originate from the polarization and the magnetization of these virtual pairs when the vacuum is stressed by an electrostatic or magnetostatic field respectively. We obtain numerical values very close to the measured values. The exact equalities constraint the free parameters of our vacuum model. 
Then we show that if we simply model the propagation of a photon in vacuum as a succession of transient captures with virtual pairs, we can derive a finite velocity of the photon with a magnitude close to the measured speed of light $c$. Again this is the occasion to adjust better our vacuum model. Since the transit time of a photon is a statistical process we expect it to be fluctuating and this translates into a fluctuation of $c$ which, if measured, would bring another piece of information on the vacuum.

When submitted to a stress the vacuum may change and this will induce a variation in the electromagnetic constants. We show this to be the case around a gravitational mass. It gives a physical interpretation of a varying vacuum refractive index equivalent to the curved space-time in General Relativity. 
The known measurements of the deflection of light by a mass, the Shapiro delay and the gravitational redshift do bring constraints on the way inertial masses should depend upon the vacuum.

At last some experimental predictions are proposed.

\end{abstract}

\section{Introduction}
Vacuum is one of the most intriguing concepts in physics. In Quantum Field Theory, the quantum vacuum is not empty and is filled with quantized fields with non zero ground energy levels.
The uncertainty principle in quantum physic leads to fluctuations of their expectation energy values which can be associated to virtual particles filling the vacuum. 
The growing role of the vacuum in fundamental phenomenon's is witnessed in the Lamb shift~\cite{lamb}, the variation of the fine structure constant with energy~\cite{bare-charge}, or the magnetic moment of the electron~\cite{magnetic-moment} and of the muon~\cite{Davier}.
All these effects are understood as due to the vacuum polarization and are well calculated in the framework of QED, or even QCD for the magnetic moment of the muon.
The vacuum can also be seen as the triggering actor in the spontaneous decay of excited atomic states through a virtual photon stimulating the emission\cite{Purcell}. In that particular effect experimentalists were able to change the vacuum, producing a huge increase or a decrease~\cite{Goy,Hulet} of the spontaneous emission rate. Along the same line of thoughts, Casimir predicted that a pressure would be present between electrically neutral conducting surfaces~\cite{Casimir}. This force has been observed in the last decade by several experiments~\cite{Casimir-exp-1} and is interpreted as arising from the modification of the zero-point energies of the vacuum due to the presence of material boundaries.

The vacuum can be stressed by several means and the effects of static electric or magnetic fields has been theorized~\cite{Latorre} and should reflect, for instance, in a modification of the speed of light $c$. What is done is the calculation of the departure from $c$ due to these vacuum modifications. The value of $c$ is assumed and is not calculated. Not only $c$ but the vacuum permittivity $\epsilon_0$, and the vacuum permeability $\mu_0$ are assumed to be fundamental constants and their experimental values have no known physical origin.
The goal of this paper is to propose a mechanism arising from the quantum vacuum which would lead naturally to these three electromagnetic constants.
While we were writing up this paper, we found an article recently published in~\cite{Leuchs} where the authors propose a similar approach to explain the physical origin of $\epsilon_0$ and $\mu_0$. Although the mechanism that they propose to derive these two constants is different from the one we will propose in this paper, the original idea is the same: {\it the existence of the physical electromagnetic constants whose numerical values are simply determined experimentally, emerge naturally from the quantum theory}. To our knowledge, we do not know of any other paper which proposes a direct derivation of $\epsilon_0$ and $\mu_0$ and we do not know any mechanism arising  from the quantum vacuum which lead to a finite photon velocity $c$. The most important consequence in our model is that $c$, $\epsilon_0$ and $\mu_0$ are not fundamental constants but are observable parameters of the quantum vacuum: they can vary if the vacuum varies in space or in time. For instance, as shown in section~\ref{sec:gravitation}, the vacuum is modified in the presence of a gravitational field, leading to a variation of $c$, $\epsilon_0$ and $\mu_0$  and a physical interpretation of a varying vacuum refractive index equivalent to the curved space-time in a gravitational field. 

The paper is organized as follows. First we describe our model of the quantum vacuum filled with virtual charged fermion pairs and we show how $\epsilon_0$ and $\mu_0$ might originate from the polarization and the magnetization of these virtual pairs. Then we show that if we simply model the propagation of the photon in vacuum as a series of interactions with virtual pairs, we can derive a velocity of the photon with a magnitude surprisingly close to the measured speed of light. After that, we study the response of the quantum vacuum under the action of a gravitational field and discuss the consequences. Finally we present a few experimental tests that could be at variance with the standard views and in particular we predict statistical fluctuations of the transit time of photons across a vacuum path, which translate into fluctuations of $c$.

%
%

\section{An effective description of quantum vacuum}
\label{sec:model}

We propose to model some electromagnetic and gravitational properties of quantum vacuum as filled with virtual pairs of elementary particles.
We consider here only virtual pairs of charged fermions.
All species of charged fermions are taken into account: the three families of charged leptons $e$, $\mu$ and $\tau$ and the three families of quarks $(u,d)$, $(c,s)$ and $(t,b)$, including their three color states.

We will take sums over the fermion type noted $i$. We use the notation $Q_i=q_i/e$, where $q_i$ is the modulus of the $i$-kind fermion electric charge and $e$ the modulus of the electron charge.

Virtual pairs of type $i$ are assumed to be produced from virtual photons filling the vacuum. Their density is assumed to be limited by the Pauli principle. This implies that when a pair decays back into virtual photons,
the infinite vacuum photon density allows to create immediately a new pair at the same place.

We assume that first order properties can be safely deduced assuming all pairs are created with an average energy\ $\mathcal{E}_i$, not taking into account a full probability density of the pair kinetic energy.
The average energy of the pair is taken proportional to its rest mass energy:
\begin{eqnarray}
\label{eq:model-1}
\mathcal{E}_i\  = K_\mathcal{E}\ 2 m_i c^2
\end{eqnarray}
where $m_i$ is the rest mass of the fermion $i$ and $K_\mathcal{E}$ an unknown constant, assumed to be independent from the fermion type, which we keep as a free parameter, greater than unity.

The life-time $\tau_i$ is given by the Heisenberg uncertainty principle:
\begin{eqnarray}
\label{eq:tau}
\tau_i\ = \frac{\hbar}{2\mathcal{E}_i}
\end{eqnarray}

The largest pair life-time is the $e^+ e^-$ one, with a value in the $ 10^{-22}\ s$ range.


We also assume that the vacuum density $N_i$ of the $i$ virtual pairs is fixed by the minimum distance $\delta_i$ allowed between two virtual fermion in the same spin state. This distance is physically limited by the Pauli exclusion principle and is of the order of half the Compton wavelength $\lambda^C_{i}$ of the fermion. We set: 
\begin{eqnarray}
\label{eq:compton-length}
\delta_i = K_\delta \frac{\lambda^C_{i}}{2} = \frac{K_\delta\hbar}{2 m_i c} 
\end{eqnarray}
where the $K_\delta$ factor is assumed to be the same for all fermion types and is expected to be of the order of unity. 
Due to angular momentum conservation, the fermion pairs are coming from the vacuum in the singlet spin state. So,
there are two fermion-antifermion spin combinations (namely up-down or down-up) per cell, and the $i$-pair density is:
\begin{eqnarray}
\label{eq:density}
N_i= \frac{2}{\delta_{i}^3}
\end{eqnarray}
With $K_\delta=1$, $\delta_i$ ranges from $200\ fm$ for $e^+e^-$ pairs to $.6\ am$ for $t\bar{t}$ pairs, and $N_i$ ranges from $ N_e \approx  2.8 \ 10^{38}$~pairs/m$^3$ to $ N_t \approx  1.5 \ 10^{50}$~pairs/m$^3$.


%
%

\section{Derivation of the vacuum permittivity}
\label{sec:permittivity}

Consider a parallel-plate capacitor with a gas inside. When the pressure of the gas decreases the capacitance decreases too until there is no more molecules in between the plates.
The strange thing is that the capacitance is not zero when we hit the vacuum. In fact the capacitance has a very sizeable value as if the vacuum were a usual material body.
The dielectric constant of a medium is coming from the existence of opposite electric charges that can be separated under the influence of an applied electric field $E$.
Furthermore the charge separation stays finite because they are bound in a molecule. These opposite translations result in no charge in the volume of the dielectric and opposite charge
appearing on the dielectric surfaces in regard of the metallic plates. This leads to a decrease of the effective charge, which implies a decrease of the voltage across the dielectric slab and finally
to an increase of the capacitance.

In our model of the vacuum the virtual pairs are the pairs of opposite charges and the separation stays finite because the electric field acts only during the life time of the pairs. In an absolute {\it empty} vacuum the induced charges would be null because there would be no charges to be separated and the capacitance of our parallel-plate capacitor would go to zero
when we would remove all molecules of the gas. 

We will see in this section that introducing our vacuum filled by virtual fermions will cause its electric charges to be separated and to appear at the level of $\epsilon_0 \approx 9\ pCb/m$ ($\approx 5.10^7$ electron charges $/m$) under an electric stress $E = 1\ V/m$.

We assume that every fermion-antifermion virtual pair of the $i$-kind bears a mean electric dipole $d_i$ given by:
\begin{eqnarray}
\label{eq:elecdipole}
\vec{d_i} = eQ_iK_d\vec{\delta_i}
\end{eqnarray}
where $\delta_i$ the distance between the pairs given by eq. \ref{eq:compton-length}, and $K_d$ an arbitrary factor assumed universal, which is expected to be of the order of unity.

If no external electric field is present, $\vec{\delta_i}$ points randomly in any direction and the resulting average field produced by the collection of these virtual dipoles is zero.


We propose to give a physical interpretation of the observed vacuum permittivity $\epsilon_0$ as originating from the mean polarization of these virtual fermions pairs in presence of an external electric field $\vec{E}$.
This polarization would show up due to the dipole life-time dependence on the electrostatic coupling energy of the dipole to the field. At distances large compared to $\delta_i$, in a field homogeneous at $\delta_i$ scale, this energy is $\vec{d_i}\cdot\vec{E}$. From (\ref{eq:tau}), one gets:

\begin{eqnarray}
\label{eq:taudipel}
\tau_i(\theta)= {{\hbar/2} \over {\mathcal{E}_i-\vec{d_i}\cdot\vec{E}}} = {{\hbar} \over {2\mathcal{E}_i({1-\eta_i \cos \theta})}}
\end{eqnarray}
$$ {\rm with}\ \theta=(\widehat{\vec{d_i},\vec{E}})\ {\rm and\ } \eta_i={{d_iE}\over{\mathcal{E}_i}}$$

Since it costs less energy to produce such an elementary dipole aligned with the field, this configuration lasts a bit longer than the others, leading to an average dipole different from zero. This average dipole $D_i$ is aligned with $\vec{E}$. Its value is obtained by integration over $\theta$ with a weight proportional to the pair life-time:
$$ D_i = {{\int_0^{\pi} d_i\ \cos\theta\  \tau_i(\theta)\ 2\pi \sin\theta\ d\theta}\over{\int_0^{\pi} \tau_i(\theta)\ 2\pi \sin\theta\ d\theta}}$$
To first order in $E$, one gets: 
\begin{eqnarray}
\label{eq:polar}
D_i = {{d_i}{\eta_i\over 3}} = {{ d_i^2}\over {3\mathcal{E}_i}} E
\end{eqnarray}

We estimate the permittivity $\tilde{\epsilon}_{0,i}$ due to $i$-type fermions using the relation
\begin{eqnarray}
\label{eq:defepsi}
P_i=\tilde{\epsilon}_{0,i}E
\end{eqnarray}
where the polarization $P_i$ is equal to the dipole density:
\begin{eqnarray}
\label{eq:defepsi}
P_i=N_iD_i
\end{eqnarray}
Thus:
\begin{eqnarray}
\label{eq:epsi}
\tilde{\epsilon}_{0,i} = {{N_i D_i}\over E}={{ 2d_i^2}\over {3\mathcal{E}_i\delta_i^3}}=\frac{1}{3}
\frac{K_d^2}{K_\mathcal{E}}{{ e^2Q_i^2}\over {m_ic^2\delta_i}}
\end{eqnarray}
So:
$$\tilde{\epsilon}_{0,i}  = {\frac{2K^2_d}{3K_\mathcal{E}K_\delta}}{e^2\over {\hbar c}}\ Q_i^2$$
which does depend only on the fermion electric charge, and no more on its mass.

Each species of fermions increases the induced polarization and therefore the vacuum permittivity. By summing over all the fermions, one gets the estimation $\tilde{\epsilon}_0$ of ${\epsilon}_0$:
\begin{eqnarray}
\label{eq:epsi0}
\tilde{\epsilon}_0 =  {\frac{K^2_d}{K_\mathcal{E}K_\delta}}\frac{2 e^2}{3 \hbar c}\sum_{i}{Q_i^2}
\end{eqnarray}

The sum is taken over all fermion types. We do not have to sum on the anti-fermions which are already included in the dipole of the virtual pair nor on the two spin states which are accounted for in the density in Equation~(\ref{eq:density}). 
For one generation $\sum_{i}{Q_i^2} = 1+3 \times(4/9+1/9)=8/3$ (the factor 3 stands for color).
Each generation contributes. Hence, for the three families of the standard model $\sum_{i}{Q_i^2} = 8$, and one obtains:
\begin{eqnarray}
\label{eq:permittivity}
\tilde{\epsilon}_0 = {\frac{K^2_d}{K_\mathcal{E}K_\delta}} \frac{16}{3} \frac{e^2}{\hbar} \frac{1}{c}
\end{eqnarray}
If $K_d \approx K_\mathcal{E} \approx K_\delta \approx 1$, one gets $\tilde{\epsilon}_0 \approx 4.3 \ 10^{-12}$~F.m$^{-1}$, a numerical value very close to the observed value $\epsilon_0 = 8.85 \ 10^{-12}$~F.m$^{-1}$. If we require an exact equality, we obtain a first relation constraining three parameters of our model of vacuum
\begin{eqnarray}
\label{eq:firstconstraint}
{\frac{K^2_d}{K_\mathcal{E}K_\delta}} =\frac{3}{64\pi\alpha}\approx 2.04
\end{eqnarray}

As in~\cite{Leuchs}, the vacuum permittivity does not depend upon the fermion masses, but only upon the number of species and the parameters of the proposed model of vacuum. This is at variance with the common idea that the energy density of the vacuum is the dominant factor~\cite{Latorre}.

\section{Derivation of the vacuum permeability}
\label{sec:permeability}
The vacuum acts as a highly paramagnetic substance. When a torus of a material is energised through a winding carrying a current I, there is a resulting magnetic flux density B which is expressed as:  
\begin{eqnarray}
\label{eq:mu-1}
B = \mu_0 n I + \mu_0 M
\end{eqnarray}
where $n$ is the number of turns per unit of length, $nI$ is the magnetic intensity in A/m, $M$ is the corresponding magnetization induced in the material and is the sum of the induced magnetic moments divided by the corresponding volume. 
In an experiment where the current $I$ is kept a constant and where we lower the quantity of matter in the torus, $B$ decreases. As we remove all matter, $B$ gets to a non zero value: $B = \mu_0 n I$ showing experimentally that the vacuum is paramagnetic with a vacuum permeability $\mu_0 = 4\pi\ 10^{-7} {N/A^2}$.

We propose to give a physical interpretation to the observed vacuum permeability as originating from the magnetization of the charged virtual fermions pairs under a magnetic stress,  following the same procedure as in the former section.

Each charged virtual fermion carries a magnetic moment proportional to the Bohr magneton:
\begin{eqnarray}
\label{eq:magneton}
\mu_i = \frac{eQ_i\hbar}{2m_i}
\end{eqnarray}
(with the notations defined in section  \ref{sec:model}).

Since the total spin of the pair is zero, and since fermion and antifermion have opposite charges, each pair carries twice the magnetic moment of one fermion. The coupling energy of a $i$-kind pair to an external magnetic field $\vec{B}$ is then $-2\vec{\mu}_i\cdot \vec{B}$.

In a similar way as in the former section (Eq.\ \ref{eq:taudipel}),  the pair life-time is:
\begin{eqnarray}
\label{eq:taumag}
\tau_i(\theta)= {{\hbar/2} \over {\mathcal{E}_i\ -2\vec{\mu}_i\cdot \vec{B}}}
\end{eqnarray}
$$ {\rm with}\ \theta=(\widehat{\vec{\mu_i},\vec{B}})$$
As in the electrostatic case, pairs with a dipole moment aligned with the field last a bit longer than anti-aligned pairs.
This leads to a non zero average magnetic moment $<\mathcal{M}_i>$ for the pair, aligned with the field and given, to first order in $B$, by:
\begin{eqnarray}
\label{eq:magnet}
<\mathcal{M}_i> = {{4\mu_i^2}\over {3\mathcal{E}_i}} B
\end{eqnarray}
The contribution $\tilde{\mu}_{0,i}$ of the $i$-type fermions to the vacuum permeability is given by $ B=\tilde{\mu}_{0,i}M_i $ or:
\begin{eqnarray}
\label{eq:defmui}
\frac{1}{\tilde{\mu}_{0,i}}=\frac{M_i}{B}
\end{eqnarray}
where $M_i$ is the volume magnetic moment:
$$M_i  = {N_i <\mathcal{M}_i>}$$
This leads to:
\begin{eqnarray}
\label{eq:valmui}
\frac{1}{\tilde{\mu}_{0,i}}={{8\mu_i^2}\over {3\mathcal{E}_i} \delta_i^3}
\end{eqnarray}
Replacing $\delta_i,\ \mathcal{E}_i$ and $\mu_i$ by their expressions, one gets :
$$\frac{1}{\tilde{\mu}_{0,i}} = \frac{8} {3K_\delta^3 K_{\mathcal{E}}} \frac{c e^2}{\hbar}\ {Q_i^2}$$
which, again, depends only on the fermion electric charge, and no more on its mass. Again this shows that the energy density of the vacuum is not the relevant parameter.

Summing the magnetic moments coming from all fermion types, allows to give for $\mu_0$ the following estimation:
\begin{eqnarray}
\label{eq:mu0}
\tilde{\mu}_0  = \frac{3K_\delta^3 K_\mathcal{E}}{8 \sum_{i}{Q_i^2}}\frac{\hbar}{c\ e^2}
\end{eqnarray}

As in the former section, $\sum_{i}{Q_i^2} = 8$, and one obtains:
\begin{eqnarray}
\label{eq:permeability}
\tilde{\mu}_0  = K_\delta^3 K_\mathcal{E} \frac{3}{64}  \frac{\hbar}{e^2} \frac{1}{c}
\end{eqnarray}
If $K_\delta \approx K_\mathcal{E} \approx 1$, one gets $\tilde{\mu}_0 \approx 6.4 \ 10^{-7}$~N.A$^{-2}$, corresponding to a numerical value very close to the observed value $\mu_0 = 4\pi \ 10^{-7}$~N.A$^{-2}$. If we require an exact equality, we obtain a second relation constraining two parameters of our model of vacuum:
\begin{eqnarray}
\label{eq:secondconstraint}
{{K^3_\delta}{K_\mathcal{E}}}=\frac{256\pi\alpha}{3}\approx 1.96
\end{eqnarray}
Using Eq.~(\ref{eq:permeability}) with Eq.~(\ref{eq:epsi0}) from former section, $\tilde{\epsilon_0}\tilde{\mu}_0$ is found not to depend upon the vacuum energy density but moreover to be independent from the number of fermion generations:
\begin{eqnarray}
\label{eq:thirdconstraint}
\tilde{\epsilon_0}\tilde{\mu}_0 =\frac{1}{c^2} \frac{K^2_dK^2_\delta}{4}
\end{eqnarray}
This implies  $K_dK_\delta=2$, which means that the pair electric dipole should be equal to the fermion charge multiplied by the fermion Compton length, whatever
the total energy of the pair and whatever the pair density. 
\vskip.25in

Equations (\ref{eq:firstconstraint}) and (\ref{eq:secondconstraint}) allow to compute $ K_\delta$ and $K_d$ from $K_\mathcal{E}$:
\begin{eqnarray}
\label{eq:Kdelta}
K_\delta=\frac{1.25}{\sqrt[3]{K_\mathcal{E}}}
\end{eqnarray}
and
\begin{eqnarray}
\label{eq:Kd}
K_d=1.6\ \sqrt[3]{K_\mathcal{E}}
\end{eqnarray}

The bound $K_\mathcal{E}>1$, which requires that the fermion pairs are produced with some kinetic energy, converts into bounds
on $K_\delta$ and $K_d$:
$$K_\delta<1.25\ {\rm and\ } K_d>1.6$$

So far, this model is self consistent and proposes a quantum origin to electromagnetic constants.

%
%

\section{Derivation of the light velocity in vacuum}
\label{sec:speedoflight}

We have shown how the virtual particle model of vacuum explains its electrostatic and magnetostatic properties. 
The Maxwell equations impose that the speed of light $c$ is thus given by the relation $c^2=1/(\epsilon_0 \mu_0)$. However no physical process has been proposed to explain the real origin of a finite velocity for a photon of null inertial mass. 
We propose in this section a mechanism which leads to a {\it finite} speed of light arising from the interaction of the photon with the virtual pairs. 
The propagation of the photon in vacuum is modeled as a series of interactions with virtual pairs present in vacuum. 
When a real photon propagates inside the vacuum, it interacts and is temporarily captured by a virtual pair during a time corresponding to the life-time $\tau$ of the virtual pair. 
As soon as the virtual pair disappears, it releases the photon to its initial energy and momentum state.
The photon continues to propagate with a {\it bare} velocity $c_0$ which is assumed to be much greater than $c$. Then it interacts again with a virtual pair and so on. 
The delay on the photon propagation produced by these successive interactions corresponds to the finite velocity of light.

The mean free path of the photon between two successive interactions is $ \Lambda = (\sigma N)^{-1} $, 
where $\sigma$ is the cross-section for the photon capture by a virtual fermion and $N$ is the numerical density of the virtual fermion pair.

The mean time for a photon to cross a length $\Lambda$ is ${\Lambda}/{c_0} +  \tau $. It corresponds to a photon velocity, noted here $v$, given by

$$ v = \frac{\Lambda}{\frac{\Lambda}{c_0}+\tau} = \frac{1}{\frac{1}{c_0} + \sigma N \tau} $$

The {\it bare} photon velocity $c_0$ is assumed to be much greater than $c$. This bare velocity corresponds to a velocity of light in an {\it empty} vacuum with no virtual particle. It is therefore natural to assume that $c_0$ is infinite. In other words, time does not flow between two successive interactions. That means that time (as defined as the duration to go from one state or one point to another one) would physically arise from the life-time of the virtual particles and the uncertainty principle.

We obtain a general expression of the photon velocity $v$ in function of three parameters of the vacuum model: 
$$ v = \frac{1}{\sigma N \tau} $$
\vskip.25in
The cross-section $\sigma$ for a real photon to interact and to be trapped by a virtual pair of fermions during its whole life-time can be estimated from the Thomson cross-section $\sigma_{Thomson}= \frac{8 \pi}{3} \alpha^2 (\lambda^{C})^2$ where $\lambda^{C}=\hbar/(m_ec)$ is the Compton wavelength of the electron.

For Thomson interaction, the factor $\alpha^2$ corresponds to the probability $\alpha$ that the photon is temporarily absorbed by the real electron times the probability $\alpha$ that the real electron releases the photon.
However, in the case of the interaction of a photon on a virtual fermion, the second $\alpha$ factor must be ignored since the photon is released with a probability equal to $1$ as soon as the virtual pair disappears. 
Thus we write $\sigma$ as:
\begin{eqnarray}
\label{eq:sigma}
\sigma_i = K_\sigma \frac{8 \pi}{3} \alpha Q_i^2(\lambda^C_{i})^2
\end{eqnarray}
with the notation given in section \ref{sec:model}, where $K_\sigma$, expected to be of the order of unity,  is assumed not to depend on the nature of the fermion.
For instance, the cross-section on one particle of a $e^+ e^-$ virtual pair is $\sigma_e \approx $~100~barns.
We notice that the cross-section does not depend on the energy of the initial photon. It implies that the vacuum is not dispersive as it is experimentally observed.

The mean free path taking now all fermion types into account is given by:
$$\frac{1}{\Lambda}=\sum_{k}{\frac{1}{\Lambda_k}}$$
So:
$$ \Lambda= \frac{1}{\sum_{k}{\sigma_k N_k}} $$

The probability to be trapped on the $i$-kind fermion for a given interaction is:
$$ p_i = \frac{\Lambda}{\Lambda_i} $$
The average photon trap duration $\tau$ per interaction, is:
$$ \tau = \sum_{i}{p_i \tau_i} = \Lambda \sum_{i}{\sigma_i N_i \tau_i} $$
This corresponds to an average photon velocity $v$ given by the general expression:
\begin{eqnarray}
\label{eq:c-1}
v = \frac{\Lambda}{\tau} = \frac{1}{\sum_{i}{\sigma_i N_i \tau_i}}
\end{eqnarray}

Using equations~(\ref{eq:tau}),~(\ref{eq:density}) and~(\ref{eq:sigma}), one obtains the formula for the average velocity of a photon in vacuum:
\begin{eqnarray}
\label{eq:v2}
v =  \frac{3K_\delta^3K_\mathcal{E}}{K_\sigma {32 \pi} \alpha \sum_{i}{Q_i^2}}\ c
\end{eqnarray}
We notice that the photon velocity depends only on the electrical charge units $Q_i$ of the virtual charged fermions present in vacuum but does not depend on their masses, nor on the vacuum energy density. Notice this is not relying on the Maxwell relation $c=1/\sqrt{\epsilon_0\mu_0}$. We mean that the Maxwell relation is not necessarily always true.

We do not sum over the two possible spins of fermions. Indeed a photon of helicity 1 (-1 respectively) can interact only with a fermion or an antifermion of helicity $-1/2$ ($+1/2$ respectively) to flip temporarily its spin to helicity $+1/2$  ($-1/2$ respectively). But we take into account the antifermion, and again $\sum_{i}{Q_i^2}=8$.

Thus if we sum the contributions from all the charged fermions, the velocity of the photon becomes:
\begin{eqnarray}
\label{eq:speedoflight}
v =   \frac{3K_\delta^3K_\mathcal{E}}{K_\sigma{256 \pi} \alpha }\ c= \frac{K_\delta^3K_\mathcal{E}}{K_\sigma}\ 0.51\ c 
\end{eqnarray}
Using equation (\ref{eq:Kdelta}), one gets:
\begin{eqnarray}
\label{eq:speedoflightsimple}
v =   \frac{1}{K_\sigma}\ c
\end{eqnarray}
So:
\begin{eqnarray}
\label{eq:Ks}
K_\sigma=1
\end{eqnarray}

It is remarkable that this simple model leads to a numerical value of the speed of light with the correct order of magnitude. 

\vskip.125in

The average speed of the photon in our medium being c, the photon propagates, on average, along the light cone. As such, the speed of the photon is independent of the inertial frame as demanded by relativity, despite the infinite photon speed between the trapping sequences.



%
%

\section{Derivation of the vacuum refractive index produced by a gravitational field}
\label{sec:gravitation}

When submitted to a stress the vacuum may change and this will induce a change in the constants. We show this to be the case around a gravitational mass and we discuss the coherence of that result with General relativity.

Einstein was the first who noticed in his own words that ``{\it the constancy of the velocity of light can be maintained only insofar as one restricts oneself to spatio-temporal regions of constant gravitational potential}''~\cite{Einstein-1912}, and later~\cite{Einstein-1920}: 
``{\it the curvature of rays of light can only take place when the velocity of propagation of light varies with position... We can only conclude that the special theory of relativity cannot claim an unlimited domain of validity; its result hold only so long as we are able to disregard the influences of gravitational fields on the phenomena (e.g. of light)}''. 
The idea that curved spacetime in a gravitational field is equivalent to an optical medium with a graded vacuum refractive index in a flat space time was first suggested by Eddington~\cite{Eddington}. 
Rosen~\cite{Rosen} has shown how to formulate general relativity within the framework of a flat metric. 
It has been also shown (for instance by Pauli~\cite{Pauli} or by Landau and Lifshitz~\cite{Landau}) that the propagation of light described in general relativity in a static gravitational field as $\delta \int g_{00}^{-1/2} \ dl = 0$ is equivalent to the Fermat principle: $\delta \int n \ ds = 0$. 
Following this analogy, Felice~\cite{Felice}, and later Evans, Nandi and Islam~\cite{Evans-1996-2} show that in many metrics of physical interest, the gravitational field can be represented as an optical medium with an effective index of refraction which varies in space.
They show that the propagation of both photons and massive particles are governed by a variational principle which involves the index of refraction and which assumes the form of Fermat's principle for the photon and Maupertuis' principle for massive particles. 
For instance, in the Schwarzschild exterior metric which applies to the spacetime around an electrically neutral static spherical mass $M$, and far from the Schwarzschild radius, the vacuum refractive index is given by
\begin{eqnarray}
\label{eq:gravit-2}
n(r) = 1+\frac{2GM}{rc_{\infty}^2} = 1+\frac{R_s}{r}
\end{eqnarray}
where $G$ is the gravitational constant, $r$ is the radial distance to the gravitational centre in a flat spacetime, $c_{\infty}$ is the velocity of light in vacuum at infinity (vacuum with no gravitational field) and $R_s=2GM/c_{\infty}^2$ is the Schwarzschild radius.  This formula is similar to the one already proposed by Eddington. 
A different calculation using the same Fermat analogy has been also performed recently giving the same result~\cite{Ye-2008}.

A direct consequence of this approach is that the speed of light $c$ is not constant anymore but varies with the radial distance $r$ as:
\begin{eqnarray}
\label{eq:gravit-3}
c(r) = \frac{c_{\infty}}{n(r)}=c_{\infty} \left( 1-\frac{R_s}{r} \right)
\end{eqnarray}

It is shown in~\cite{Evans-1996-2} and \cite{Puthoff} that this new formulation leads naturally to the phenomenae predicted by general relativity, namely the deflection of light rays by a mass, the Shapiro delay and the gravitational redshift.  


\vskip.125in

To our knowledge, no physical interpretation has ever been proposed to explain the origin of that refractive index.
We propose here to show that our model of interaction of light with the quantum vacuum gives a possible physical origin of a varying refractive index in presence of a gravitational field.
We study the simplest case of a spherical static gravitational field produced by a mass $M$ and we give a possible mechanism leading to the right expression of the refractive index as given in Equation~(\ref{eq:gravit-2}). 


As shown in section~\ref{sec:speedoflight}, the speed of light $c(r)$ at any distance $r$ from the gravitational mass can be written as 
$c(r) = 1 / \left( N(r) \tau(r) \sigma(r) \right )$, 
where $N(r)$, $\tau(r)$ and $\sigma(r)$ are the density of virtual pairs, their life-time and their cross-section for photons capture respectively.
Infinitely far from the gravitational mass $M$, the speed of light is 
$c_{\infty} = 1 / \left( N_{\infty} \tau_{\infty} \sigma_{\infty} \right )$. 
The vacuum refractive index $n(r)$ is thus defined by the relation
\begin{eqnarray}
\label{eq:gravit-7}
n(r) = \frac{c_{\infty}}{c(r)} = \frac{N(r)\sigma(r)\tau(r)}{N_{\infty} \sigma_{\infty} \tau_{\infty}}
\end{eqnarray}

Before calculating how $N(r)$, $\tau(r)$ and $\sigma(r)$ should vary when the mass $M$ stresses the vacuum, we must first make some hypothesis on which fundamental constants are really constant or can vary since we allow $c$ to vary.
We assume that the only fundamental constants in nature which do not vary in a gravitational field are the Planck constant $\hbar$, the electric charge $e$ and the gravitational constant $G$. 
If $\hbar$ and $e$ are constant then $\alpha$ appears to be also constant in our model. Indeed Equation~(\ref{eq:permittivity}) shows that $\epsilon_0$ varies as $ e^2/(\hbar c)$. The product $\epsilon_0.c$ is thus constant, and $\alpha = e^2/(4 \pi \epsilon_0 \hbar c)$ is constant.
However the mass energy $mc^2$  and the inertial mass $m$ vary. The way the mass energy varies must be consistent with the experimental observations of the effect of a gravitational potential difference on the apparent energy of an atom. This gravitational redshift effect has been initially observed with $^{57}$Fe by Pound and Rebka~\cite{PoundRebka-1960}~\cite{PoundRebka-1964} and confirmed later with high precision with a hydrogen-maser~\cite{Vessot}. A possible physical interpretation of this effect within our formalism is the following. The energy $h\nu$ of the emitted photon is constant when it is going closer to the Earth, but its velocity $c$ (and therefore its wavelength) varies. The atomic energy level, which is proportional to the Rydberg constant $R_y=\alpha^2/2 \times m_ec^2$ is smaller when the atom is closer to the Earth due to the variation of $c$ and consequently of the mass energy. To be consistent with the experimental observations, the mass energy $mc^2$ must vary as
\begin{eqnarray}
\label{eq:gravit-7bis}
m(r)c^2(r) = m_{\infty} c^2_{\infty} - \frac{GMm_{\infty}}{r} =  m_{\infty} c^2_{\infty} \left( 1-\frac{R_s}{2r} \right)
\end{eqnarray}
In other words, the mass energy of a particle at a distance $r$ from a gravitational mass is equal to its mass energy without gravitation field plus the gravitational potential. Comparing this relation with Equation~(\ref{eq:gravit-3}), it appears that the mass energy $mc^2$ varies as $c^{1/2}$ and consequently the mass $m$ as $c^{-3/2}$. This behaviour is in agreement with pionneer models developped initially by Wilson~\cite{Wilson} and later by Dicke~\cite{Dicke} who proposed a theory where gravitation could be explained as a result of a varying vacuum permittivity near matter. In their formalism, the rest energy $mc^2$ must vary as $m_{\infty} c^2_{\infty}/\sqrt{\epsilon_0}$. But, as shown in section~\ref{sec:permittivity} Equation~(\ref{eq:epsi0}), our model leads to a vacuum permittivity which varies as $c^{-1}$. Thus the relation proposed by Wilson is in agreement with a mass energy which varies as $c^{1/2}$ or an inertial mass which varies as $c^{-3/2}$.

The consequence of a varying mass energy is that the life-time $\tau$ of the virtual pairs is modified by a gravitational field. Without field, equation (\ref{eq:tau}) can be rewritten as: 
\begin{eqnarray}
\label{eq:gravit-8bis}
\tau_{\infty}=\frac{\hbar/2}{K_{\mathcal{E}}2m_{\infty}c_{\infty}^2}
\end{eqnarray}
In the presence of a static spherical gravitational field, the average energy of the virtual pair depends upon the gravitational potential energy (see Equation~(\ref{eq:gravit-7bis})) and its life-time becomes:
\begin{eqnarray}
\label{eq:gravit-8}
\tau(r) = \frac{\hbar/2}{K_{\mathcal{E}}2m(r)c^2(r)} = \frac{\hbar/2}{K_{\mathcal{E}}2m_{\infty}c_{\infty}^2 (1- \frac{R_s}{2r})}= \tau_{\infty} \left(1-\frac{R_s}{2r}\right)^{-1}
\end{eqnarray}
Far from the Schwarzschild radius, the Equation~(\ref{eq:gravit-8}) may be approximated by:
\begin{eqnarray}
\label{eq:gravit-10}
\tau(r) \approx  \tau_{\infty} \left(1 + \frac{R_s}{2r} \right)
\end{eqnarray}
It means that the life-time of the virtual pairs is larger when we get closer to a gravitational mass $M$.

If gravitation acts in the same direction for fermions and antifermions, the virtual pair of charged fermions behaves under a gravitational field as a monopole. Therefore the virtual pair should move towards the mass $M$ during its life-time $\tau$. 
Suppose a unit volume $dS.dr$ where $dS$ is the unit area perpendicular to the distance $r$ from the gravitational mass M. The mass of virtual pairs contained in the unit volume is $N(r) dS dr 2m$ and the gravitational force on the unit volume is thus given by:
\begin{eqnarray}
\label{eq:gravit-11}
F(r) = \frac{GM}{r^2}N(r) dS dr 2m(r)
\end{eqnarray}
But the virtual pair composed of fermions cannot move freely in space. The presence of other virtual fermions in its neighbouring pushes them back due to the Pauli exclusion principle which forbids any overlaping of fermionic wave functions with the same quantum states. Such an effect corresponds to an effective Pauli pressure which opposes the gravitational force. A pressure is proportional to an energy density, thus the Pauli pressure should be proportional to the energy density $N(r)mc^2$ of the virtual fermions (or antifermions):
\begin{eqnarray}
\label{eq:gravit-12}
p_{Pauli} = \eta 2 N(r) m(r)c^2(r) 
\end{eqnarray}
where $\eta$ is a proportional constant which must be close to 1.\\
At equilibrium, the difference of the Pauli pressure forces on each side of the unit volume at any distance $r$ is equal to the gravitational force. It corresponds to
\begin{eqnarray}
\label{eq:gravit-13}
2 \eta \left( N(r+dr)-N(r) \right) m(r)c^2(r) dS = - \frac{GM}{r^2}N(r) 2 m(r) dS dr 
\end{eqnarray}

We obtain the differential Equation for the virtual pair density:
\begin{eqnarray}
\label{eq:gravit-14}
\frac{dN}{N(r)} = - \frac{1}{\eta} \frac{GM}{r^2 c(r)^2} dr
\end{eqnarray}
Far from the Schwarzschild radius, the integration leads to first order in $R_s/r$ to
\begin{eqnarray}
\label{eq:gravit-15}
N(r) = N_{\infty} exp\left( + \frac{1}{\eta} \frac{GM}{rc_{\infty}^2}\right) \approx  N_{\infty} \left(1 + \frac{1}{\eta} \frac{R_s}{2r} \right)
\end{eqnarray}

It means that the density of virtual pairs is larger when we get closer to a gravitational mass $M$. Consequently the occupancy size $\delta$ of a virtual pair is smaller near a mass. From Equation~(\ref{eq:density}), $\delta$ varies as
\begin{eqnarray}
\label{eq:gravit-15bis}
\delta(r) = \delta_{\infty} \left( \frac{N(r)}{N_{\infty}} \right)^{-1/3} = \delta_{\infty} \left(1 - \frac{1}{3\eta} \frac{R_s}{2r} \right)
\end{eqnarray}

We can now calculate how the cross-section $\sigma(r)$ of photon capture by a virtual pair varies along $r$ in the gravitation field. 
From Equations~(\ref{eq:compton-length}) and (\ref{eq:sigma}) the cross-section is proportional to $\sigma(r) \sim \alpha \delta^2$. However, as discussed before, $\alpha$ appears to be constant in a gravitational field. Therefore $\sigma(r)$ varies as $\delta^2$
\begin{eqnarray}
\label{eq:gravit-18}
\sigma(r) = \sigma_{\infty} \left(\frac{\delta(r)}{\delta_{\infty}} \right)^2 = \sigma_{\infty} \left(1 - \frac{1}{3\eta} \frac{R_s}{r} \right)
\end{eqnarray}

The combination of Equations~(\ref{eq:gravit-7}), (\ref{eq:gravit-10}), (\ref{eq:gravit-15}) and (\ref{eq:gravit-18}) leads to the expression of the vacuum refractive index:
\begin{eqnarray}
\label{eq:gravit-19}
n(r) = \left(1 + \frac{R_s}{2r} \right) \left(1 + \frac{1}{\eta}\frac{R_s}{2r} \right) \left(1 - \frac{2}{3\eta}\frac{R_s}{2r} \right)
\end{eqnarray}

With $\eta=1/3$, one obtains the right expression of the vacuum refractive index as derived from general relativity and given in Equation~(\ref{eq:gravit-2}).

\vskip.125in

Since the light velocity is not constant in this formalism but depends on the gravitation potential, we should expect an annual modulation of $c$ due to the eccentricity of the Earth orbit. The annual variation of the distance $R$ of the Earth to the Sun is $\Delta R/R \approx 0.03$. Thus the expected annual variation of $c$ is 
$$\frac{\Delta c}{c} = \frac{2GM_{\odot}}{R c_{\infty}^2} \frac{\Delta R}{R} \approx 6 \ 10^{-10} $$
where $M_{\odot}$ is the mass of the sun. 

Experimental test of local Lorentz invariance and possible variation of fundamental constants have been recently performed with high sensitivity by comparing the resonance frequency $\nu_{cso}$ of a cryogenic sapphire oscillator (CSO) with a reference frequency $\nu_{H}$ produced by a hydrogen maser. Results of a long-term operation over a period of greater than six years~\cite{Tobar-Wolf} show no annual modulation with a sensitivity of the order of $(\nu_{cso}-\nu_{H})/\nu_{H} < 10^{-14}$/day.
This result is in agreement with our model of varying light velocity.  
Indeed the CSO resonant frequency $\nu_{cso}$ is proportional to the ratio of $c$ over the length of the sapphire crystal $L$: $\nu_{cso} \propto c/L$. 
As $c$ varies, the Bohr radius $R_B = 4 \pi \epsilon_0 \hbar^2/(m_e e^2)$ and consequently $L$ vary as $c^{1/2}$. Thus $\nu_{cso}$ varies as $c^{1/2}$. The reference maser frequency $\nu_{H}$ is refered  to an atomic energy level. Thus $\nu_{H}$ varies as the Rydberg constant, which varies as $c^{1/2}$ as discussed before. Thus the ratio $(\nu_{cso}-\nu_{H})/\nu_{H}$ appears constant in our model.


%
%

\section{Observational predictions}
\label{sec:predictions}
\subsection{c variation with $\vec{E}$ and $\vec{B}$}
This model allows to make a simple prediction on vacuum properties due to the $e^+e^-$ pair life-time sensitivity to $\vec{E}$ and $\vec{B}$ fields. As shown in sections~\ref{sec:permittivity} and \ref{sec:permeability}  the pair life-times depend on the mean polarization induced by external fields. Eq. (\ref{eq:taudipel}) and (\ref{eq:taumag}) can be written as:
\begin{eqnarray}
\label{eq:tautou}
<\tau_i>(E,B)= {{\hbar/2} \over {\mathcal{E}_i-<D_i> E-<\mathcal{M}_i>B}} 
\end{eqnarray}
since the electric dipoles align on $\vec{E}$ and the spins on $\vec{B}$ independently.
This leads to:
\begin{eqnarray}
\label{eq:tauquad}
<\tau_i>(E,B)=\tau_i(0,0) {{1} \over {1-\frac{1}{3}\left(\frac{d_i E}{\mathcal{E}_i}\right)^2-\frac{1}{3}\left(\frac{2\mu_i B}{\mathcal{E}_i}\right)^2}} 
\end{eqnarray}
which gives, as a function of model parameters
\begin{eqnarray}
\label{eq:taudevel}
<\tau_i>(E,B)=\tau_i(0,0) {{1} \over {1-\frac{1}{3}\left(\frac{eQ_i\lambda^C_i E}{K_\mathcal{E}2m_ic^2}\right)^2-\frac{1}{3}\left(\frac{eQ_i\lambda^C_i c B}{K_\mathcal{E}2m_ic^2}\right)^2}} 
\end{eqnarray}
Setting
\begin{eqnarray}
\label{eq:ecrit}
E_i^c=\frac{K_\mathcal{E}2m_i^2c^3} {eQ_i\hbar}
\end{eqnarray}
one gets simply:
\begin{eqnarray}
\label{eq:tauecrit}
<\tau_i>(E,B)=\tau_i(0,0) {{1} \over {1-\frac{1}{3}\left(\frac{E}{E^c_i}\right)^2-\frac{1}{3}\left(\frac{c B}{E^c_i}\right)^2}} 
\end{eqnarray}
The energy of the electric or magnetic dipoles in fields  $E_i^c$ or $E_i^c/c$ are equal to the fermion pair mass (including kinetic) energy. These fields have the smallest values for electrons: respectively $$E^c_e=2.6\ 10^{18} K_\mathcal{E}\ V/m$$ and $$B^c_e=8.8\ 10^9K_\mathcal{E}\ T$$
Propagating this life-time change into the light velocity formula (\ref{eq:c-1}), and taking only the electron effect into account, leads to an effective refraction index in presence of static fields:
\begin{eqnarray}
\label{eq:effec}
n(E,B)=\frac{c(0,0)}{c(E,B)}=\frac{7}{8}+ \frac{1}{8}\ {{1} \over {1-\frac{1}{3}\left(\frac{E}{E^c_e}\right)^2-\frac{1}{3}\left(\frac{B}{B^c_e}\right)^2}} 
\end{eqnarray}
These effects are expected to show up whatever the orientation of the fields with respect to photon direction. Its numerical value depends on $K_{\mathcal{E}}$ and is:
$$n(E,B)-1=\frac{6\ 10^{-27} E^2(MV/m)+ 5\ 10^{-22}B^2(T)}{K^2_{\mathcal{E}}}$$

This prediction can be tested in high field laboratories. For instance, a forthcoming experiment~\cite{Pelle} planning to check magneto-electrical properties of quantum vacuum as computed from QED~\cite{Rizzo} and aiming at sensitivities in the $10^{-26}$ range on indexes, should make use of fields as high as $20\ MV/m$ and $15\ T$, for which we predict a $(2\ 10^{-24}+10^{-20})/{K^2_{\mathcal{E}}}$ effect.


\subsection{Transit time fluctuation}

Quantum gravity theories including stochastic fluctuations of the metric or compactified dimensions, predict a fluctuation $\sigma_t$ of the propagation time of photons~\cite{Yu-Ford}. However  observable effects are expected to be too small to be experimentally tested. 
Hogan\cite{Hogan} also predicts that the non commutative geometry at the Planck scale should produce a Planckian noise corresponding to a spatially coherent uncertainty in transverse rest frame velocity of photon. Hogan and collaborators are developing an interferometer experiment to test it. Ellis et al. also predict that vacuum  should be dispersive meaning that photons of different energies would propagate at different velocities~\cite{ellis-2000}.
The observation of gamma ray bursts (GRB) has been used to place limits on such a possible variation of $c$~\cite{Abdo}.

In our model we also expect fluctuations of the speed of light $c$. Indeed in the mechanism proposed here $c$ is due to the effect of successive interactions and transient captures of the photon with the virtual particles in the vacuum. Thus statistical fluctuations of $c$ are expected, due to the statistical fluctuations of the number of interactions $N_{stop}$ of the photon with the virtual fermions and the life-time fluctuation of the virtual pairs.

The propagation time of a photon which crosses a distance $L$ of vacuum is
$$ t = \sum_{i,k}{ t_{i,k}} $$
where $t_{i,k}$ is the duration of the $k^{th}$ interaction on an $i$-kind fermion. Let $ N_{stop,i}$ be the mean number of such interactions.
The variance of $t$ due to the statistical fluctuations of $N_{stop,i}$ is:
$$ \sigma_{t,N}^2 = \sum_{i}{N_{stop,i} \tau_i^2} $$
The life-time should also fluctuate. Assuming exponential probability distributions of average values $\tau_i$, the variance of the sum of the independent $t_{i,k}$ is:
$$ \sigma_{t,\tau}^2 = \sum_{i}{N_{stop,i} \tau_i^2} $$
So
\begin{eqnarray}
\label{eq:fluctu}
\sigma_t^2 =2 \sum_{i}{N_{stop,i} \tau_i^2}
\end{eqnarray}
Since $ N_{stop,i}=L/\Lambda_i$, one has:
\begin{eqnarray}
\label{eq:fluctuation-0}
\sigma_t^2 = 2 L \sum_{i}{\frac{\tau_i^2}{\Lambda_i}} =2 L \sum_{i}{\sigma_i N_i \tau_i^2} 
\end{eqnarray}
Using equations~(\ref{eq:tau}), (\ref{eq:density}) and (\ref{eq:sigma}), it comes:
$$ \sigma_t^2 = \frac{K_\sigma}{K_\mathcal{E}^2K_\delta^3}\frac{16\pi\alpha}{3}\frac{\hbar}{c^3} \sum_{i}{\frac{Q_i^2}{m_i}} \times L $$ 
The contribution of each fermion is inversely proportional to its mass. Therefore the fluctuations of the propagation time are dominated by virtual $e^+ \ e^-$ pairs. Summing only on the electrons and positrons, $\sum_{i}{Q_i^2/m_i}=1/m_e$ since only one fermion spin state couples to a given photon helicity, and the anti-fermion is accounted for in the density definition. One obtains:
\begin{eqnarray}
\label{eq:fluctuation-1}
\sigma_t = \sigma_0 \sqrt{L}
\end{eqnarray}
with:
\begin{eqnarray}
\label{eq:fluctuation-2}
\sigma_0  = 4 \sqrt{\frac{K_\sigma}{K_\mathcal{E}^2K_\delta^3} \frac{\pi \alpha\hbar}{3\ m_ec^2\ c}}\approx  \sqrt{\frac{K_\sigma}{K_\mathcal{E}^2K_\delta^3} }\ 0.7 \ fs\ m^{-1/2}
\end{eqnarray}
Replacing $K_\delta$ and  $K_\sigma$ by expressions (\ref{eq:Kdelta}) and (\ref{eq:Ks}) allows to give $\sigma_0$ as a function of $K_\mathcal{E}$:
\begin{eqnarray}
\label{eq:sigmazero}
\sigma_0  \approx \frac{.5}{ \sqrt{K_\mathcal{E} } }\ fs\ m^{-1/2}
\end{eqnarray}

Our model predicts a fluctuation of $c$ and a measurement of this effect would deliver precious information on the vacuum.

The experimental way to test fluctuations is to measure a possible time broadening of a light pulse travelling a distance $L$ of vacuum. The figure of merit to constrain this model is $\sqrt{L} / \Delta t$ where $\Delta t$ is the time width of the light pulse. 
The very fact that these statistical fluctuations should go like the square root of the distance implies the exciting idea that experiments on Earth do compete with astrophysical constraints as we will see now.
\vskip.125in

\noindent{\bf{Constraints from astrophysical observations}}

\noindent The very bright GRB 090510, detected by the Fermi Gamma-ray Space Telescope~\cite{Abdo}, at MeV and GeV energy scale,
present short spikes in the 8~keV - 5~MeV energy range, with the narrowest widths of the order of 10ms. Observation of optical after glow, a few days later by ground based spectroscopic telescopes give a common redshift of $z = 0.9$. This corresponds to a distance, in the standard cosmology, of about $2\ 10^{26} m$. Translated into our model this is saying that the fluctuation of $c$ has to be smaller than about $0.7 fs\ m^{-1/2}$. It is important to notice that there is no expected dispersion of the bursts in the interstellar medium at this energy scale. 

If we move six orders of magnitude down in distances we arrive to kpc and pulsars.
Short microbursts contained in main pulses from the Crab pulsar have been recently observed at the Arecibo Observatory telescope at 5 GHz~\cite{Crab-pulsar-2010}. The frequency-dependent delay caused by dispersive propagation through the interstellar plasma is corrected using a coherent dispersion removal technique. 
The mean time width of these microbursts after dedispersion is about 1~$\mu$s, much larger than the expected broadening caused by interstellar scattering. If this new unknown broadening is correlated to the emission properties, this implies fluctuations of $c$ smaller than about $0.2 fs\ m^{-1/2}$.

In these observations of the Crab pulsar, some very sporadic pulses with a duration of less than $1 ns$ were observed at 9 GHz~\cite{Crab-pulsar-2007}.
This is 3 orders of magnitude smaller than the usual pulses. These nanoshots can occasionally be extremely intense, exceeding $2 MJy$, and have
an unresolved duration of less than $0.4\ ns$. The light-travel size $c\delta t \approx 12 cm$. From this the implied brightness temperature is $2\ 10^{41} K$.
Alternatively we might assume the emitting structure is moving outward with Lorentz factor  $\gamma_b \approx  10^2 - 10^3$. In that case, the size estimate
increases to $10^3 - 10^5 cm$, and the brightness temperature decreases to $10^{35} - 10^{37} K$. 
We recall that the Compton temperature is $10^{12} K$ and that the Planck temperature is $10^{32} K$ so the phenomenon, if real, would be
way beyond known physics. Another way to look at the extraordinary physics behind such events is to consider the energy which has to
be liberated. It is equivalent to the total annihilation of 100 tons of matter in a beach ball volume and in less than a nanosecond.
We emphasis also two features. Firstly, these nanoshots are contained in a single time bin (2~ns at 5~GHz and 0.4~ns at 9~GHz) corresponding to a time width less than $2/\sqrt{12} \approx 0.6 $ns at 5~GHz and $0.4/\sqrt{12} \approx 0.1$ns at 9~GHz, below the expected broadening caused by interstellar scattering. Secondly, their frequency distribution appear to be almost monoenergetic. 
However, if these nanoshots are really true, the fluctuations of the vacuum would be a thousand times smaller. 

Going down a further $10^{17}$ in distances, which is going from the kpc to a few hundred meters, we expect fluctuations in the $fs$ range.
Therefore an experimental setup using femtosecond lasers sent to a multi-pass cavity should be able to detect such a phenomenon and help characterizing the vacuum.
If nanoshots from the Crab pulsar are true, we would have to move from femtosecond to attosecond laser techniques.

\subsection{Annual modulation of $\epsilon_0$ and $\mu_0$}

We have shown in section~\ref{sec:gravitation} that the speed of light $c$ varies when  photons move towards a mass $M$ (eq.~\ref{eq:gravit-3}). Thus Equations~(\ref{eq:epsi0}) and~(\ref{eq:mu0}) show that both $\epsilon_0$ and $\mu_0$ should increase while we go towards a gravitational mass $M$, as $c^{-1}$:
$$ \epsilon_0 \approx 10 \frac{e^2}{\hbar} \frac{1}{c_{\infty}} \left( 1+\frac{2GM}{r c_{\infty}^2} \right) $$
$$ \mu_0 \approx 0.1 \frac{\hbar}{e^2} \frac{1}{c_{\infty}} \left( 1+\frac{2GM}{r c_{\infty}^2} \right) $$
An annual modulation of $\epsilon_0$ or $\mu_0$ is therefore expected with a relative variation due to the excentricity of the Earth orbit given by
$$ \frac{\Delta \epsilon_0}{\epsilon_0} = \frac{\Delta \mu_0}{\mu_0} = \frac{2GM_{\odot}}{R c_{\infty}^2} \frac{\Delta R}{R}$$
where $M_{\odot}$ is the mass of the sun, R is the average distance of the Earth to the Sunwith a seasonal variation of the order of $\Delta R/R \approx 0.03$.  Thus the expected annual modulation is
$ (\Delta \epsilon_0)/\epsilon_0 = (\Delta \mu_0)/\mu_0 \approx 6 \ 10^{-10} $.


The seasonal variation of $\epsilon_0$ can be illustrated by the variation of the capacity $C$ of two metallic plates of surface $S$ separated by a distance $d$. 
The capacity is given by $C=\epsilon_0 S/d$. 
Since $c$ varies, the Bohr radius $R_B = 4 \pi \epsilon_0 \hbar^2/(m_e e^2)$, and consequently the size of the atoms vary as $c^{1/2}$. Thus the distance $d$ varies as $c^{1/2}$ and the surface $S$ as $c$. Then the capacity $C$ varies as $c^{-1/2}$ corresponding to an annual relative modulation of $3~10^{-10}$.

\subsection{mumesic atoms}
This model predicts short distance deviations to standard electrostatics at scales much larger than for the QED screening effects in the one photon exchange process.

Section \ref{sec:permittivity} is based on eq.~(\ref{eq:taudipel}) valid only at large distance compared to the dipole size.
The dipole size being found to be equal to the fermion Compton length, one expects a decrease of $\epsilon_0$ up to $12.5\%$ for distances small compared to $\lambda^C_e\approx 400 fm$, since at such scales the $e^+e^-$ pairs do not contribute to the polarization.

In mumesic Hydrogen, the muon Bohr radius is about $250 fm$ and it has been observed an increase in the energy of $5\%$~\cite{Pohl}. This result has been interpreted as a need for shrinking the proton size, but it corresponds to a $2.5\%$ decrease of $\epsilon_0$ which is a natural consequence of this model which is suited to describe bound states. We predict that this effects is higher in mumesic higher $Z$ atoms for which measurements are planned.

\section{Conclusions}

We proposed a quantum virtual particles model of the vacuum including the concept of density and  life time of virtual particles. Within this framework, we have shown how $\epsilon_0$ and $\mu_0$ originate from the polarization and the magnetization of these virtual pairs when the vacuum is stressed by an electrostatic or a magnetostatic field respectively. We proposed that the finite speed of light could be due to successive transient captures of photon with virtual particle present in the vacuum. Our calculated values for $\epsilon_0$, $\mu_0$ and $c$ are very close to the measured values. The exact equalities constrain the free parameters of our vacuum model. 
In these proposed model, the vacuum may change when submitted to a stress and this will induce a change in the electromagnetic constants. We showed this to be the case around a mass. The gravitational stress upon the vacuum induces a change of $c$ which is reflected into the well known angular deflection of the photons, the Shapiro delay and the gravitational redshift.
This physical interpretation of a varying vacuum refractive index analogy to General Relativity do bring constraints on the way inertial masses should depend upon the vacuum.

This model predicts several effects:

Stressing the vacuum with an electric or a magnetic field implies a change in the speed of light in the reach of soon forthcoming experiments.

The propagation of a photon being a statistical process we expect a fluctuation of the speed of light. It is shown that this could be within the grasp of nowadays experimental techniques and we plan to assemble such an experiment.

Mumesic atoms energy levels probe this model at short distance.





\end{document}